\documentclass[letterpaper,aps,prl,twocolumn]{revtex4}
\usepackage{graphicx,bm}
\usepackage{amsmath,amssymb}

\begin{document}

\title{Pairing mechanism in multiband superconductors}
\author{Wen-Min Huang and Hsiu-Hau Lin}
\affiliation{Department of Physics, National Tsing Hua University,
300 Hsinchu, Taiwan}
\affiliation{Physics Division, National Center for Theoretical Sciences, Hsinchu 300, Taiwan}
\date{\today}
\begin{abstract}
It has been a long-standing puzzle why electrons form pairs in unconventional superconductors, where the mutual interactions are repulsive in nature. Here we find an analytic solution for renormalization group analysis in multiband superconductors, which agrees with the numerical results exceedingly well. The analytic solution allows us to construct soluble effective theory and answers the pairing puzzle: electrons form pairs resonating between different bands to compensate the energy penalty for bring them together, just like the resonating chemical bonds in benzene. The analytic solutions allow us to explain the peculiar features of critical temperatures, spin fluctuations in unconventional superconductors and can be generalized to cuprates where the notion of multiband is replaced by multipatch in momentum space. Therefore, finding effective attractions between electrons is no longer a necessity and the secret for higher superconducting temperatures lies in boosting pair hopping between different bands.
\end{abstract}
\maketitle

It has been 100 years since Heike Kamerlingh Onnes discovered the resistance of mercury suddenly drops to zero\cite{Onnes1911} when cooling down by liquid helium in 1911 and marked the birth of superconductivity. The exotic phenomena of superconductors remained mysterious until Bardeen, Cooper and Schrieffer (BCS)\cite{BCS1957} came up with a complete theory in 1957. Despite the celebrating success of BCS theory, there are other superconductors remain queer\cite{Norman2011} and cannot be explained solely by the electron-phonon interactions, including cuprates\cite{Bednorz1986,Wu1987}, heavey-fermion compounds\cite{Steglich1979,Stewart1984,Joynt2002}, organic superconductors and recently found iron-based superconductors\cite{Hosono2008,Stewart2011}. Perhaps the most intriguing puzzle for these unconventional superconductors is the pairing mechanism: what is the glue to pair up electrons from mutual repulsive interactions? The experimental evidences suggest that pairing in the unconventional superconductors is not due to electron-phonon interactions. Due to strong magnetic correlations\cite{Cruz2008,Huang2008,Zhao2008,Inosov2010} in these materials, it is proposed that spin fluctuations\cite{Miyake1986,Emery1986,Scalapino86,Maier2008,Mazin2008} may play the role of glue to pair electrons up. Recent renormalization-group (RG) studies\cite{Wang2011} indeed reveals the close relation between spin fluctuations and unconventional superconductivity in iron-based materials.

In this Letter, we investigate the pairing mechanism in multiband superconductors by the unbiased RG approach. In general, the RG equations are coupled nonlinear first-order differential equations and make any simple understanding beneath the messy numerics inaccessible. However, making use of classification scheme by RG exponents, we show that the dominant interactions are intra-band $g$ and inter-band $g_\perp$ Cooper scattering. It is rather surprising that these two dominant couplings are captured by a set of analytic solutions. The solutions are elegant and simple enough to reveal the pairing mechanism in multi band superconductors.

The binding energy for Cooper pair formation is dictated by a small parameter $\delta = g - |g_\perp|$. As long as the inter-band pair hopping is larger than the intra-band, Cooper pairs resonating between different bands become stable despite of the repulsive intra-band interaction. The picture of resonating Cooper pairs between different bands leads to several natural consequences. First of all, though the coupling strengths of $g$ and $g_\perp$ are large, the binding energy of Cooper pairs (and thus the critical temperature) is determined by their difference $\delta$, which is one order smaller than the bare couplings as detailed later. Secondly, it can be shown that the inter-band pair hopping also give rise to spin fluctuations at the nesting vector which connects the dominant Fermi surfaces. The inter-band pair hopping not only explains why strong magnetic fluctuations often show up in unconventional superconductors but also pins down at what momentum the spin fluctuations should appear.

To see how the analytic solutions emerge from the RG analysis, we choose the iron-based superconductor as an demonstrating example.We start with a five-orbital tight-binding model\cite{Kuroki2008,Wang2009,Wang2011} for iron-based superconductors with generalized on-site interactions,
\begin{eqnarray}\label{model}
H &=& \sum_{\bm{p},a,b}\sum_{\alpha}
c^{\dag}_{\bm{p}a \alpha}K_{ab}(\bm{p})c_{\bm{p}b\alpha}
+ U_1\sum_{i,a}n_{ia\uparrow}n_{ia\downarrow}
\nonumber\\
&+&U_2\sum_{i,a<b}\sum_{\alpha,\beta}n_{ia\alpha}n_{ib\beta}
+J_H\sum_{i,a<b}\sum_{\alpha,\beta}
c^{\dag}_{ia\alpha}c_{ib\alpha}c^{\dag}_{ib\beta}c_{ia\beta}
\nonumber\\
&+&J_H\sum_{i,a<b} \left[ c^{\dag}_{ia\uparrow}c^{\dag}_{ia\downarrow}
c_{ib\downarrow} c_{ib\uparrow}+{\rm H.c.}\right] \bigg\},
\end{eqnarray}
where $a,b=1,2,...,5$ label the five $d$-orbitals of Fe, $1:d_{3Z^2-R^2}$, $2:d_{XZ}$, $3:d_{YZ}$, $4:d_{X^2-Y^2}$, $5:d_{XY}$. The kinetic matrix $K_{ab}$ in the momentum space has been constructed in previous studies\cite{Kuroki2008}. The generalized on-site interactions consist of three parts: intra-orbital $U_1$, inter-orbital $U_2$ and Hund's coupling $J_H$. To simplify the RG analysis, we sample each pocket with one pair of Fermi points (required by time-reversal symmetry). This is equivalent to a four-leg ladder geometry with quantized momenta and the effective Hamiltonian consists of five pairs of chiral fermions as shown in Fig. 1.

\begin{figure}
\centering
\includegraphics[width=8.5cm]{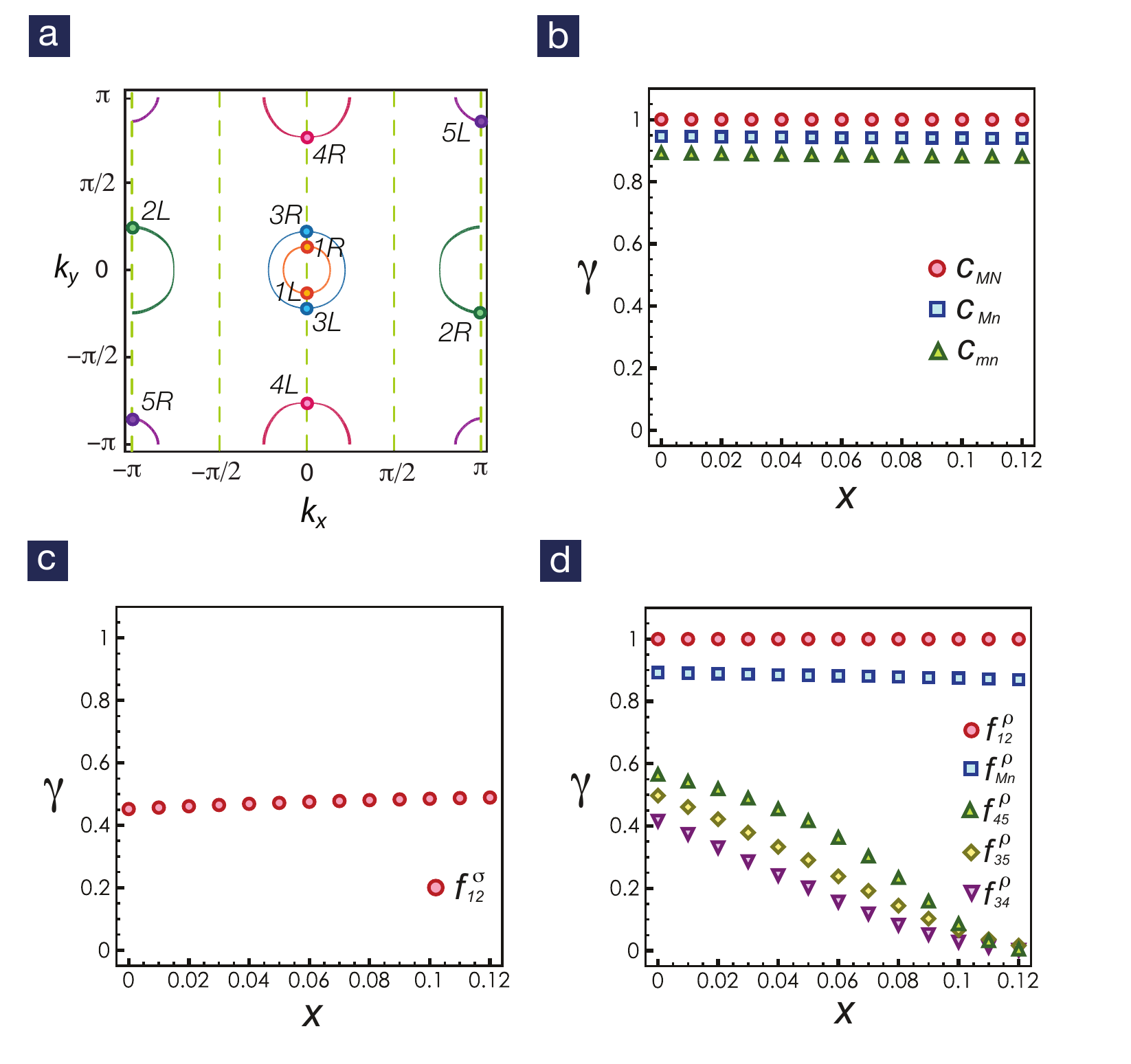}
\caption{(a) Fermi surfaces for the five-orbital Hamiltonian at electron doping $x=0.1$ with ladder geometry. The RG exponents for (b) Cooper scattering (c) forward scattering in spin sector (d) forward scattering in charge sector are presented. The dominant bands are $M,N=1,2$, while the subdominant ones are $m,n=3,4,5$. The interaction profile is set to $U_1=2.8$ eV, $U_2=1.4$ eV and $J_H=0.7$ eV.
 }
\label{FS}
\end{figure}

In weak coupling\cite{Balents1996,Lin1997,Lin1998}, the allowed interactions are Cooper scattering $c^{\sigma}_{ij}, c^{\rho}_{ij}$ and forward scattering $f^{\sigma}_{ij}, f^{\rho}_{ij}$ between different bands, where $\sigma, \rho$ denote the spin and charge channels respectively. We integrate the coupled RG equations numerically and find all couplings are captured by the scaling Ansatz\cite{Lin2005,Shih2010},
\begin{eqnarray}
g_i(l) \approx \frac{G_i}{(l-l_d)^{\gamma_i}},
\end{eqnarray}
where $G_i$ is an order one constant and $0 \leq \gamma_i \leq 1$. The scaling exponent $\gamma_i$ help us to build the hierarchy of all relevant couplings without ambiguity. 

In the doping range $x=0$ to $x=0.12$, these exponents are shown in Fig. 1. The couplings with $\gamma_i=1$ are the most dominant, including intra-band Cooper scattering $c^{\rho/\sigma}_{11}, c^{\rho/\sigma}_{22}<0$ and the inter-band ones $c^{\rho/\sigma}_{12}>0$ between the hole pocket centered at $(0,0)$ (band 1) and the electron pocket at $(\pm \pi, 0)$ (band 2). The positive sign of the inter-band Cooper scattering $c_{12}$ leads to the $s_{\pm}$ pairing symmetry. Note that our RG Ansatz predicts the dominant superconducting bands with the correct pairing symmetry as obtained by the fRG method. In fact, a detail analysis including all subdominant relevant couplings lead to correct signs for all gap functions in different bands. 

Introduce the couplings, 
$c \equiv (c^\rho_{11}+c^\rho_{22}+c^\sigma_{11}+c^\sigma_{22})/4$
for intraband pair hopping and 
$c_\perp \equiv (c^\rho_{12}+c^\rho_{21}+c^\sigma_{12}+c^\sigma_{21})/4$
for interband pair hopping. It is remarkable that the RG flows obtained in numerics, as shown in Fig. 2, are well captured by the analytic solutions $g(l)$ and $g_\perp(l)$,
\begin{eqnarray}
c(l) &\approx& g(l) = \frac{1}{2}\left(\frac{1}{l-l_+}+\frac{1}{l-l_-}\right),
\nonumber\\
c_\perp(l) &\approx& g_{\perp}(l)=\frac{1}{2}\left(\frac{1}{l-l_+}-\frac{1}{l-l_-}\right), 
\label{solution}
\end{eqnarray}
where $l_{\pm}=-1/\left[g(0)\pm g_{\bot}(0)\right]$ can be extracted numerically. Because the RG flows diverge at $l=l_d$, it fixes $l_{-}=l_d$. Meanwhile, it is reasonable to require that $c(0)+c_\perp(0) = g(0)+g_\perp(0)$, ensuring the coupling strengths are of the same order. Thus, the two parameters $l_{\pm}$ can be determined without any fitting.

\begin{figure}
\centering
\includegraphics[width=7.5cm]{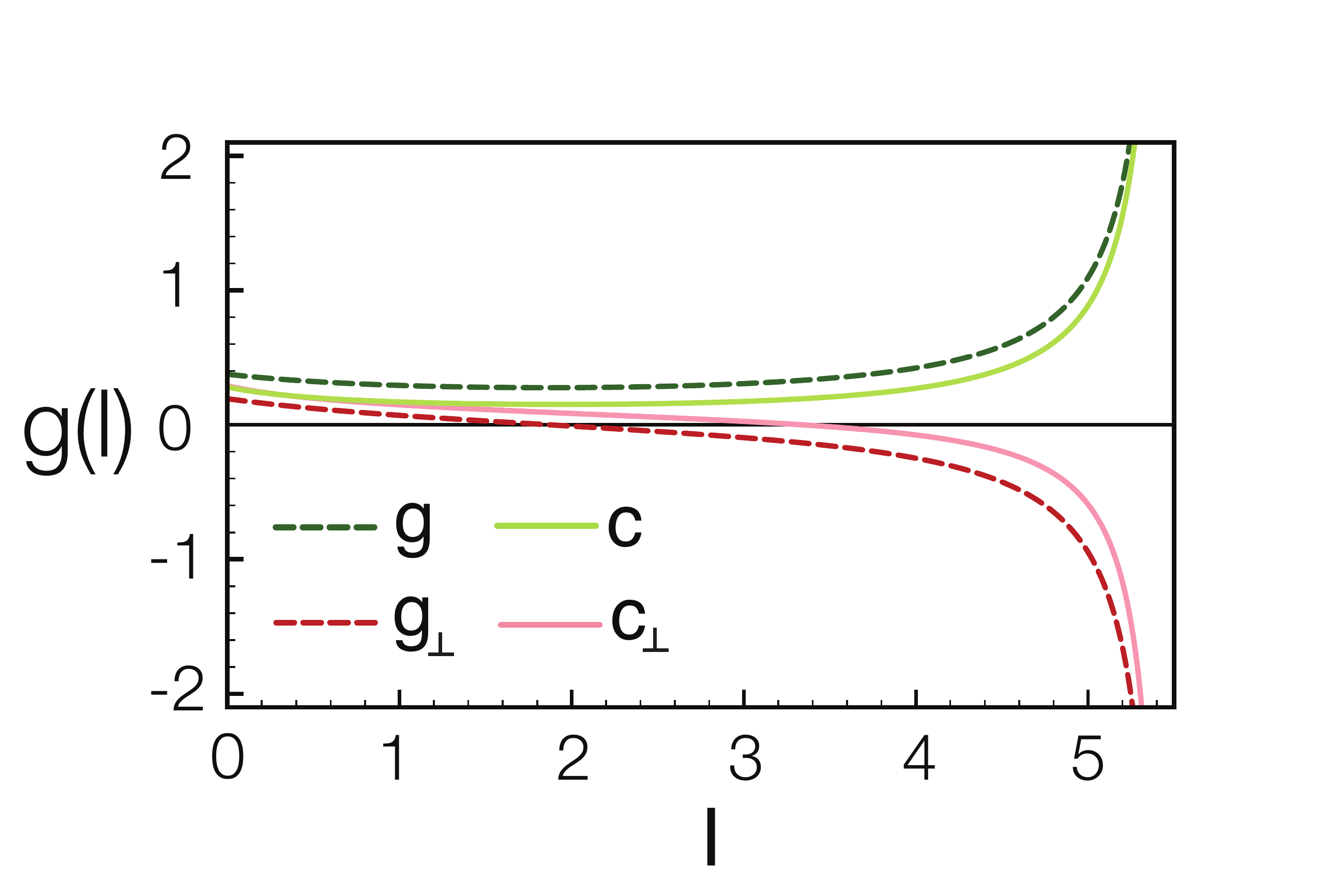}
\caption{Comparison between the analytic solutions for $g(l)$ and $g_\perp(l)$ and the numerical RG flows for $c(l)$ and $c_\perp(l)$.}
\label{MF}
\end{figure}

What is the secret message behind the analytic solutions? Consider a two-band BCS Hamiltonian with intra-band $g$ and inter-band $g_\perp$ pair hopping. Though not widely known, RG equations for these couplings can be derived from the dependence of the gap functions. After some algebra, the RG equations for $g$ and $g_\perp$ are
\begin{eqnarray}
\frac{dg}{dl}&=&-g^2-g_{\perp}^2,
\nonumber\\
\frac{dg_\perp}{dl}&=&-2gg_{\perp}.
\label{twoRG}
\end{eqnarray}
The above non-linear coupled equations can be solved exactly, giving the analytic solutions we discussed previously. Therefore, the effective theory for iron-based superconductors is the multiband BCS Hamiltonian, proven by matching the RG flows together.

\begin{figure}
\centering
\includegraphics[width=7.5cm]{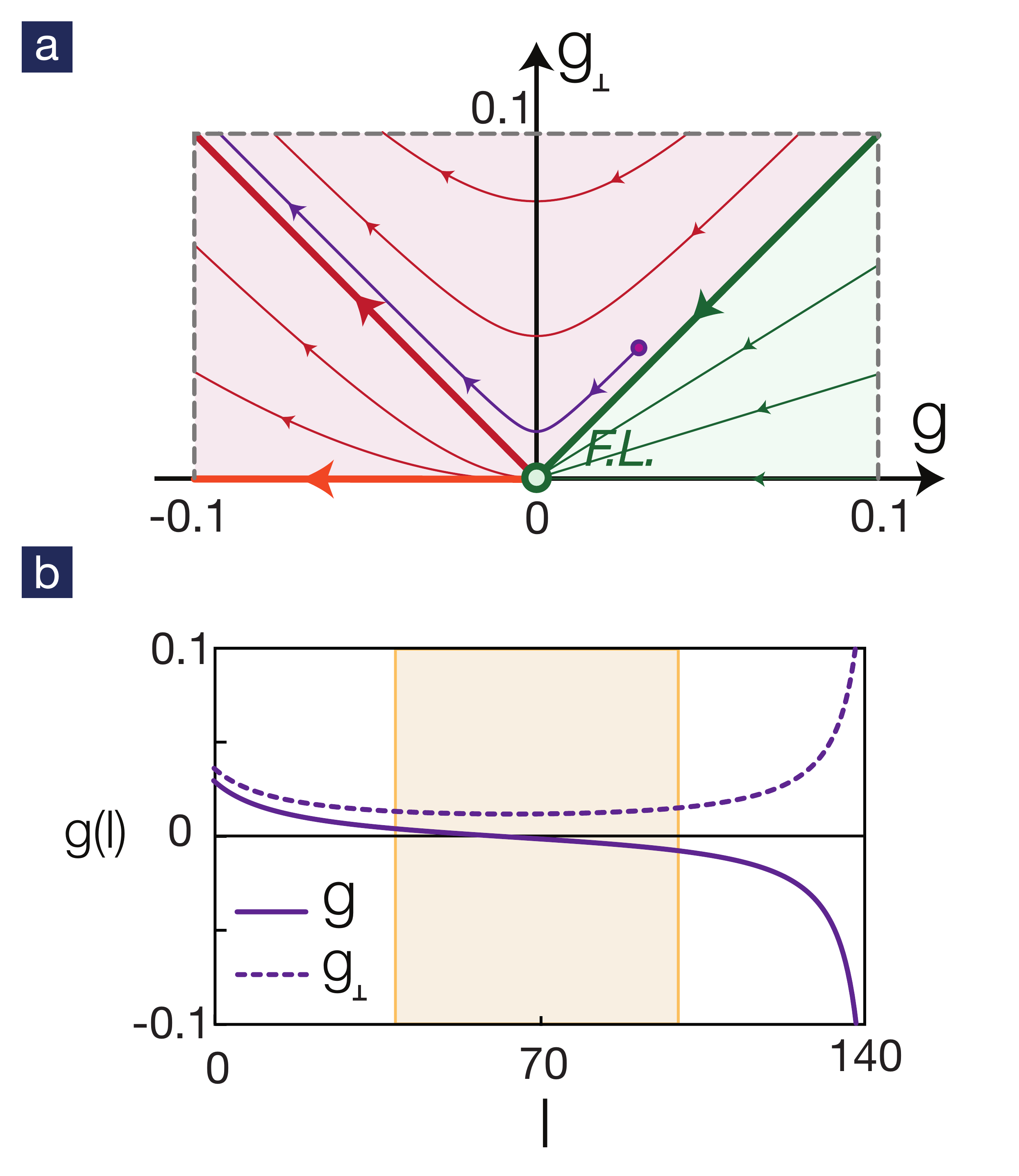}
\caption{(a) RG flows for two-band BCS Hamiltonian hosting three different phases: Fermi liquid, conventional and unconventional $s_{\pm}$-wave superconductors. (b) Typical RG flows for bare couplings close to the symmetric ray $g_\perp=g$. The corresponding trajectory on the $g-g_\perp$ plane is also illustrated in part (a).}
\label{RGplane}
\end{figure}

\begin{figure}
\centering
\includegraphics[width=8cm]{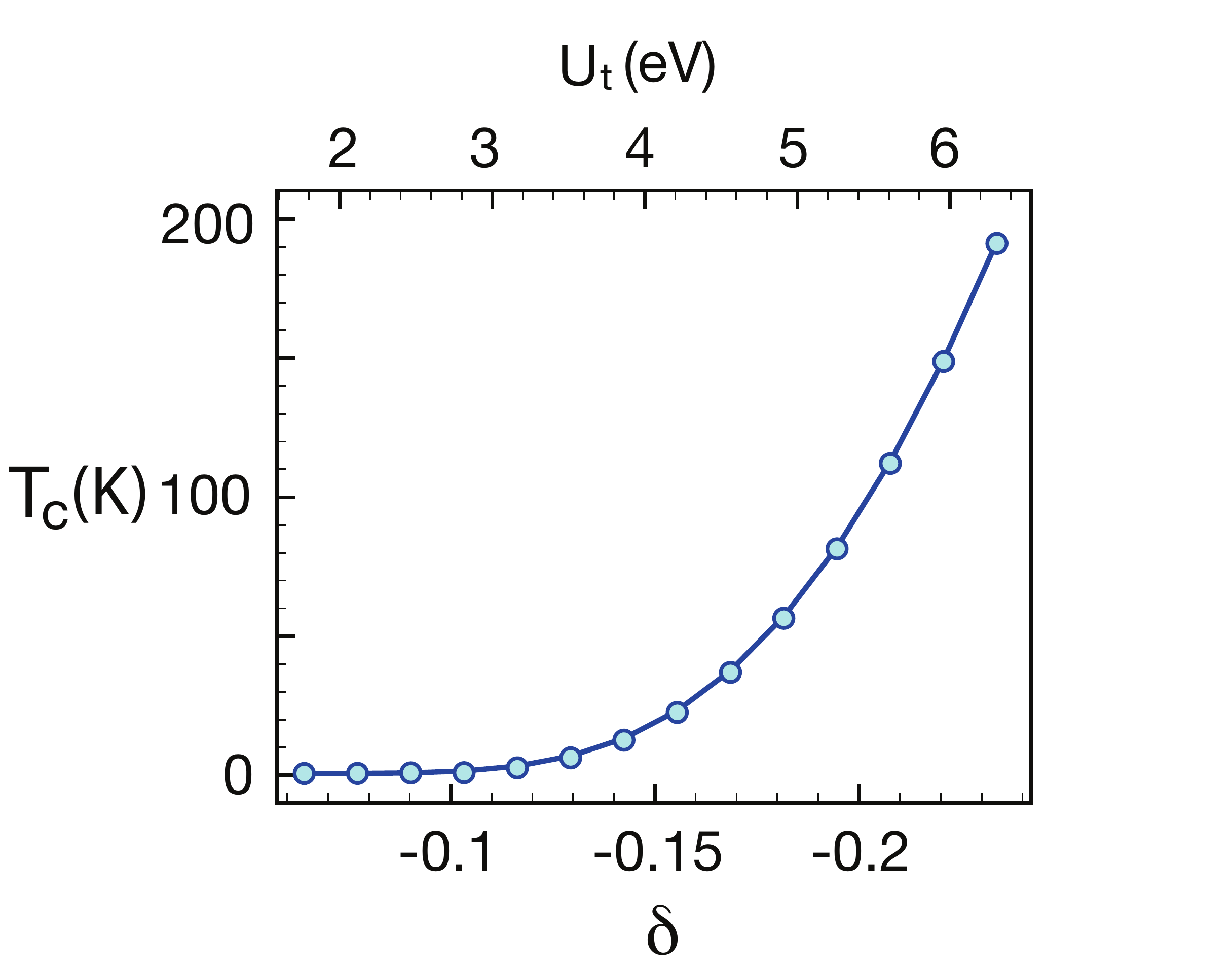}
\caption{Critical temperature $T_c$ versus the interaction strength. The on-site interaction strength is $U_{t}=U_1+U_2+J_H$ with the interaction profile $U_1/U_2=2$ and $U_1/J_H=4$. The critical temperature is dictated by the small parameter $\delta$, which can be extracted from numerical RG flows and is one order smaller than the bare interaction strength $U_t$.}
\label{Tc}
\end{figure}

RG flows for $g$ and $g_\perp$ are shown in Fig. 3. If the intra-band pair hopping is larger than the inter-band one ($g-g_\perp>0$), the couplings flow toward the Fermi-liquid fixed point. If the inter-band pair hopping is larger ($g-g_\perp<0$), they flow toward the superconducting phase with $s_\pm$ pairing symmetry. It is worth mentioning that, if the initial couplings are close to the symmetric ray $g=g_{\perp}$, it will flow toward the Fermi-liquid fixed point first and then turns around to the unconventional superconducting state. This means that, upon cooling down the system toward the critical temperature, it exhibits non-trivial crossover properties over a wide range of temperatures as shown in Fig. 3(b). However, in the absence of inter-band pair hopping ($g_\perp=0$), a negative $g(0)$, i.e. attractive interaction, is required to trigger the superconducting instability. The RG flows in the special case $g_\perp=0$ give rise to the commonly accepted criterion that an effective attraction is necessary for Cooper pair formation. However, the criterion is obviously wrong as the RG flows in the upper plane is quite different from those in the horizontal axis.

The key parameter for Cooper pair formation in multiband superconductors is $\delta = g - |g_\perp|$. Despite of the energy penalty $g$ to form a Cooper pair within the same band, through inter-band pair hopping, a Cooper pair gains $-|g_\perp|$ benefit through resonating between different bands, just like the resonating chemical bonds in benzene. Thus, attractive interactions are no longer necessary. Following the textbook calculations, the critical temperature is
\begin{eqnarray}
k_BT_c\simeq1.14~\Lambda ~e^{-1/|\delta|},
\end{eqnarray}
where $\Lambda$ is of the same order of electronic band width. Setting $\Lambda=t=1eV$, reasonable estimate for the hopping amplitude $t$, we extract $\delta$ from the numerical RG flows for different on-site interaction strengths $U_{t}=U_1+U_2+J_H$ and plot the critical temperatures in Fig. 4. The parameter $\delta$ is one order of magnitude smaller that the bare coupling $U_t/t$. If one chooses $U_1=2.8$, $U_2=1.4$ and $J_H=0.7$, the predicted critical temperature is about $56$ K -- quite a reasonable estimate.

The inter-band pair hopping $g_\perp$ brings in another interesting property in unconventional superconductors. Making use of the instability analysis developed by Wang and Lee\cite{Wang2011}, the interband pair hopping also enhances spin-density-wave (SDW) instability if the momentum $\bm{Q} = \bm{K}_1-\bm{K}_2$, connecting the two Fermi surfaces, is close to half of the reciprocal lattice vectors, i.e. $\bm{Q} = \bm{G}/2$. In the case studied here, $\bm{Q}=(\pi,0)$ satisfies the above condition. Therefore, it is expected that the antiferromagnetic spin fluctuations at the momentum $\bm{Q}$ are enhanced along with unconventional superconductivity.

Generalizing Shankar's seminal work\cite{Shankar1994} in two dimensions, we consider a two-pocket model with generic 4-fermion interactions including intra-band forward scattering $F_{PP}(\theta_1,\theta_2)$, inter-band forward scattering $F_{P\bar{P}}(\theta_1,\theta_2)$, intra-band Cooper scattering $C_{PP}(\theta_1,\theta_2)$ and inter-band Cooper scattering $C_{P\bar{P}}(\theta_1,\theta_2)$, where $P=1,2$ is the band index with the convention $(\bar{1}, \bar{2})=(2, 1)$ and $\theta_i$ represents the angle of the corresponding momentum. Detail derivations of the RG equations will be given elsewhere. In the absence of inter-band pair hopping, we have checked that our RG equations reduce to those derived by Shankar. Under RG transformation, the forward scattering does not flow but the Cooper scattering is described by the RG equations, 
\begin{widetext}

\begin{eqnarray}
\dot{C}_{PP}(\theta_1,\theta_2)&=&-\int_{0}^{2\pi}\frac{d\theta}{2\pi}
\left[
C_{PP}(\theta_1,\theta)C_{PP}(\theta,\theta_2)
+C_{P\bar{P}}(\theta_1,\theta)C_{P\bar{P}}(\theta,\theta_2)
\right],
\nonumber\\
\dot{C}_{P\bar{P}}(\theta_1,\theta_2)&=&-\int_{0}^{2\pi}\frac{d\theta}{2\pi}\left[
C_{PP}(\theta_1,\theta)C_{P\bar{P}}(\theta,\theta_2)
+C_{P\bar{P}}^l(\theta_1,\theta)C_{\bar{P}\bar{P}}(\theta,\theta_2)\right].
\end{eqnarray}
\end{widetext}
Assuming the density of states and the bare couplings are rotationally invariant, i.e. $G(\theta_1,\theta_2)=G(\theta_1-\theta_2)$, the RG equations can be decoupled by the partial-wave decomposition. By identifying $C_{PP} \to g$ and $C_{P\bar{P}} \to g_\perp$, the same set of RG equations as in Eq.~\ref{twoRG} appears and leads to similar unconventional superconductivity in two dimensions. 

If both Fermi surfaces in the two-pocket model locate at $\bf{K}_i=0$, the superconductivity is still driven by the inter-band pair hopping but it does not come with any enhanced antiferromagnetic spin fluctuations. To some extent, one can say the spin fluctuations are ``symptoms" of unconventional superconductivity but not cause of it. It would be truly exciting to look for realistic multiband superconductors with the band structure discussed here: unconventional superconductivity without enhanced antiferromagnetic spin fluctuations.

Our previous calculations concentrate on the dominant two bands. But, it is reasonable to include all active bands in the generalized BCS Hamiltonian when quantitative accuracy is required. In addition, for single-band materials but with significant variations in density of states at different momenta, the pairing mechanism discussed here is also at work. We have carried out primitive RG analysis on cuprates by cutting the single-band Fermi surface into 16 patches (minimum to differentiate nodal and antinodal regimes with four-fold dihedral symmetry). The dominant patches locate at $(\pi, 0)$ and $(0, \pi)$ with sign-reversed gap functions, agreeing with the $d$-wave symmetry and the enhanced spin fluctuations at $(\pi,\pi)$. How to rigorously generalize the analysis presented here from multiband superconductors to multi-patch single-band system remains open. Though the analytic solutions show that the couplings flow toward the multiband BCS Hamiltonian, which can be treated easily at mean-field level, the strong-coupling theory may not be of the same origin. However, the elegant and simple analytic solutions extracted here may provide some helpful hints in constructing the appropriate model or even the ground-state wave function.

We acknowledge supports from the National Science Council in Taiwan through grant NSC-100-2112-M-007-017-MY3. Financial supports and friendly environment provided by the National Center for Theoretical Sciences in Taiwan are also greatly appreciated.


\begin{thebibliography}{99}


\bibitem{Onnes1911}
H. Kamerlingh Onnes,
Commun. Phys. Lab. Univ. Leiden 120b, 122b, 124c (1911).

\bibitem{BCS1957}
J. Bardeen, L. N. Cooper and J. R. Schrieffer,
Phys. Rev. \textbf{108}, 1175 (1957).

\bibitem{Norman2011}
M. R. Norman,
Science \textbf{332}, 196 (2011) and references therein.

\bibitem{Bednorz1986}
J. G. Bednorz and K. A. M\"uller,
Z. Phys. B \textbf{64}, 189 (1986).

\bibitem{Wu1987}
M. K. Wu \textit{et al.},
Phys. Rev. Lett. \textbf{58}, 908(1987).

\bibitem{Steglich1979}
F. Steglich \textit{et al.},
Phys. Rev. Lett. \textbf{43},1892 (1979).

\bibitem{Stewart1984}
G. R. Stewart,
Rev. Mod. Phys. \textbf{56}, 755 (1984).

\bibitem{Joynt2002}
R. Joynt and L. Taillefer,
Rev. Mod. Phys. \textbf{74}, 235 (2002).

\bibitem{Hosono2008}
Y. Kamihara, T. Watanabe, M. Hirano and H. Hosono,
J. Am. Chem. Soc. \textbf{130}, 3296 (2008).

\bibitem{Stewart2011}
G. Stewart,
Rev. Mod. Phys. \textbf{83}, 1589 (2011).

\bibitem{Cruz2008}
C. de la Cruz \textit{et al.},
Nature \textbf{453}, 899 (2008). 

\bibitem{Huang2008}
Q. Huang \textit{et al.},
Phys. Rev. Lett. \textbf{101}, 257003 (2008).

\bibitem{Zhao2008}
J. Zhao \textit{et al.},
Nature Mater. \textbf{7}, 953 (2008).

\bibitem{Inosov2010}
D. S. Inosov \textit{et al.},
Nature Phys. \textbf{6}, 178 (2010).

\bibitem{Miyake1986}
K. Miyake, S. Schmitt-Rink and C. M. Varma,
Phys. Rev. B \textbf{34}, 6554 (1986).

\bibitem{Emery1986}
M. T. B\'eal-Monod, C. Bourbonnais, C and V. J. Emery,
Phys. Rev. B \textbf{34}, 7716 (1986).

\bibitem{Scalapino86}
D. J. Scalapino, E. Loh and J. E. Hirsch,
Phys. Rev. B \textbf{34}, 8190 (1986).

\bibitem{Maier2008}
T. A. Maier, D. Poilblanc and D. J. Scalapino,
Phys. Rev. Lett. \textbf{100}, 237001 (2008).

\bibitem{Mazin2008}
I. I. Mazin, D. J. Singh, M. D. Johannes and M. H. Du,
Phys. Rev. Lett. \textbf{101}, 057003 (2008).

\bibitem{Kuroki2008}
K. Kuroki \textit{et. al.},
Phys. Rev. Lett. \textbf{101}, 087004 (2008).

\bibitem{Wang2009}
F. Wang, H. Zhai, Y. Ran, A. Vishwanath and D.-H. Lee,
Phys. Rev. Lett. \textbf{102}, 047005 (2009).

\bibitem{Wang2011}
F. Wang and D.-H. Lee,
Science \textbf{332}, 200 (2011) and references therein.

\bibitem{Balents1996}
L. Balents and M. P. A. Fisher,
Phys. Rev. B \textbf{53}, 12133 (1996). 

\bibitem{Lin1997}
H.-H. Lin, L. Balents and M. P. A. Fisher,
Phys. Rev. B \textbf{56}, 6569 (1997).

\bibitem{Lin1998}
H.-H. Lin, L. Balents and M. P. A. Fisher,
Phys. Rev. B \textbf{58}, 1794 (1998).

\bibitem{Lin2005}
M.-H. Chang, W. Chen and H.-H. Lin,
Prog. Theor. Phys. Suppl. \textbf{160}, 79 (2005).

\bibitem{Shih2010}
H.-Y. Shih, W.-M. Huang, S.-B. Hsu and H.-H. Lin,
Phys. Rev. B \textbf{81}, 121107(R) (2010).

\bibitem{Shul}
R. Shul, B. T. Mathias and L. R. Walker,
Phys. Rev. Lett. \textbf{3}, 552 (1959).

\bibitem{Shankar1994}
R. Shankar,
Rev. Mod. Phys. \textbf{66}, 129 (1994).

\end{thebibliography}
\end{document}